\documentclass{aa}  

\usepackage{graphicx}
\usepackage{color}
\usepackage{txfonts}
\usepackage{tablefootnote}

\usepackage[version=4]{mhchem}

\makeatletter

\newcommand{\Rmnum}[1]{\expandafter\@slowromancap\romannumeral #1@}
\makeatother

\begin{document}

   \title{First Detection of Hydrogen in the $\beta$ Pictoris Gas Disk}

   %\subtitle{I. Overviewing the $\kappa$-mechanism}

   \author{P. A. Wilson\inst{1,2}
          \and
          A. Lecavelier des Etangs\inst{1,2}
          \and
          A. Vidal-Madjar\inst{1,2}
          \and
          V. Bourrier\inst{3}
          \and
          G. H{\'e}brard\inst{1,2}
          \and
          F. Kiefer\inst{4}
          \and
          H. Beust\inst{5,6}          
          \and
          R. Ferlet\inst{1,2}
          \and
          A.-M. Lagrange\inst{5,6}
          }

\offprints{paul.wilson@iap.fr}

\institute{
CNRS, UMR 7095, Institut d'Astrophysique de Paris, $98^{\mathrm{bis}}$ Boulevard Arago, F-75014 Paris, France
\email{paul.wilson@iap.fr}
\and UPMC Univ. Paris 6, UMR 7095, Institut d'Astrophysique de Paris, $98^{\mathrm{bis}}$ boulevard Arago, F-75014 Paris, France% 2
\and Observatoire de Gen\`eve, Universit\'e de Gen\`eve, 51 Chemin des Maillettes, 1290 Sauverny, Switzerland %3
\and School of Physics and Astronomy, Tel Aviv University, Tel Aviv 69978, Israel%4
\and Univ. Grenoble Alpes, Institut de Plan\'etologie et d'Astrophysique de Grenoble (IPAG, UMR  5274), F-38000 Grenoble, France%5
\and CNRS, Institut de Plan\'etologie et d'Astrophysique de Grenoble (IPAG, UMR 5274), F-38000 Grenoble, France%6
}

   \date{Received X; Accepted X}

% \abstract{}{}{}{}{} 
% 5 {} token are mandatory

  \abstract
   {%Context
   The young and nearby star $\beta$\,Pictoris ($\beta$\,Pic) is surrounded by a debris disk composed of dust and gas known to host a myriad evaporating exocomets, planetesimals and at least one planet. At an edge-on inclination, as seen from Earth, this system is ideal for debris disk studies providing an excellent opportunity to use absorption spectroscopy to study the planet forming environment. Using the Cosmic Origins Spectrograph (COS) instrument on the {\it{Hubble Space Telescope (HST)}} we observe the most abundant element in the disk, hydrogen, through the H\,\Rmnum{1} Lyman\,$\alpha$ (Ly-$\alpha$) line. We present a new technique to decrease the contamination of the Ly-$\alpha$ line by geocoronal airglow in COS spectra. This Airglow Virtual Motion (AVM) technique allows us to shift the Ly-$\alpha$ line of the astrophysical target away from the contaminating airglow emission revealing more of the astrophysical line profile. This new AVM technique, together with subtraction of an airglow emission map,  allows us to analyse the shape of the $\beta$\,Pic Ly-$\alpha$ emission line profile and from it, calculate the column density of neutral hydrogen surrounding $\beta$\,Pic. The column density of hydrogen in the $\beta$\,Pic stable gas disk at the stellar radial velocity is measured to be $\log(N_{\mathrm{H}}/1\,\mathrm{cm}^2) \ll 18.5$. The Ly-$\alpha$ emission line profile is found to be asymmetric and  we propose that this is caused by H\,\Rmnum{1} falling in towards the star with a bulk radial velocity of $41\pm6$\,km/s relative to $\beta$\,Pic and a column density of $\log(N_{\mathrm{H}}/1\,\mathrm{cm}^2) = 18.6\pm0.1$. The high column density of hydrogen relative to the hydrogen content of CI chondrite meteorites indicates that the bulk of the hydrogen gas does not come from the dust in the disk. This column density reveals a hydrogen abundance much lower than solar, which excludes the possibility that the detected hydrogen could be a remnant of the protoplanetary disk or gas expelled by the star. We hypothesise that the hydrogen gas observed falling towards the star arises from the dissociation of water originating from evaporating exocomets.} %The large abundance difference between hydrogen and oxygen in the gas disk also shows that exocomets could be the dominant suppliers of the hydrogen gas, only if the lifetimes of neutral oxygen and hydrogen differ significantly.} 


   \maketitle

%________________________________________________________________

\section{Introduction}
The $\beta$\,Pic system is a young planetary system embedded in a debris disk which is about 23 million years old \citep{mamajek14}. The debris disk which surrounds the star is composed of gas and dust and is continuously being replenished by colliding and evaporating planetesimals and exocomets \citep{ferlet_1987,lecavelier96,artymowicz97}. Due to the edge-on inclination of the system as seen from Earth, the light from $\beta$\,Pic is absorbed by the stable gas component of the circumstellar (CS) debris disk \citep{hobbs85,vidal-madjar_1986}. The light from $\beta$\,Pic is also absorbed by any orbiting exocomets which may transit the star \citep{kiefer14} as well as the interstellar medium (ISM) as the light makes it journey towards Earth \citep{lallement95}.

Previous studies of the debris disk surrounding $\beta$\,Pic have detected the presence of several atomic and molecular species using the {\it{Far Ultraviolet Spectroscopic Explorer}} ({\it{FUSE}}) \citep[e.g.][]{lecavelier2001,roberge06} and the {\it{Hubble Space Telescope}} ({\it{HST}}) \citep[e.g.][]{vidal-madjar_1994,lagrange95,jolly98,roberge00}. The origin of the gas disk surrounding $\beta$\,Pic is currently not well constrained. It is not thought to originate from the star-forming nebula due to the detected presence of CO \citep{vidal-madjar_1994}, which is a molecule that quickly dissociates \citep{vanDishoeck88}. The gas in the disk is therefore thought to be subject to a replenishment mechanism. A number of gas replenishing mechanisms have been proposed which include the photon-stimulated desorption from circumstellar dust grains \citep{chen2007}, vaporisation of colliding dust in the disk \citep{czechowski07} and the vaporisation of large bodies, such as exocomets, orbiting \citep{lecavelier96} or falling in towards $\beta$\,Pic \citep{beust89, lagrange95}.

The radiation pressure from the star is expected to greatly exceed the gravitational force causing the gas to be blown away completely. This is, however, not the case as observations have shown through the detection of gas absorption and emission lines at a similar radial velocity to $\beta$\,Pic ($\sim20.5$km/s, \citealt{hobbs85,brandeker04}) that the gas disk is in Keplerian motion around the star. This suggests a braking mechanism capable of slowing down the expansion of the gas. The gas which is subject to the radiative force of $\beta$\,Pic will largely be ionised. With each ion dynamically coupled to the other through Coulomb collisions the ionised gas behaves like a fluid causing a slow down of the gas expansion. Ionised carbon in the form of C$^+$ seen in C\,\Rmnum{2} lines is an effective braking agent and could be responsible for counteracting the radiation pressure provided it is abundant enough \citep{fernandez06}. Indeed, FUSE observations have shown an overabundance of carbon and oxygen \citep{roberge06,brandeker11} relative to refractory elements like Fe and Si. The carbon overabundance was also inferred from HARPS observations done with the 3.6\,m telescope \citep{brandeker11}. Herchel/HIFI observations of ionised carbon on the other hand argue in favour of preferential depletion as an explanation for the overabundance of carbon and oxygen rather than super solar abundance, suggesting the presence of another type of braking gas \citep{cataldi14}. Ionised gas may not supply a sufficient amount of gas braking, suggesting the presence of other braking agents such as the gas colliding with dust grains and neutral gas \citep{lagrange98, fernandez06, brandeker11}.

Hydrogen is one of the most difficult gasses to detect being mostly neutral and its ground state \citep{hobbs85} with transitions lines in the far-UV requiring spaced based observations. The absorption by the ISM also makes it challenging separating the absorption lines resulting from the absorption of light from the gas disk and the ISM. An attempt at constraining the abundance of atomic hydrogen was done by \cite{freudling95} who conducted radio observations of the 21\,cm emission line. No H\,\Rmnum{1} was detected, yet upper limits on the column density as a function of beam size was modelled ranging from 2 to $5 \times 10^{19}$\,cm$^{-2}$. \cite{lecavelier2001} obtained a far-ultraviolet spectrum of $\beta$\,Pic using the Far Ultraviolet Spectroscopic Explorer (FUSE) aimed at detecting H$_2$ absorption lines but detected none. They inferred an upper limit on the molecular Hydrogen column density at $N(\mathrm{H}_2 \leq 10^{18}$\,cm$^{-2})$. These upper limits showed that hydrogen gas is likely not the braking agent as the required mass of hydrogen would have to be higher \citep{brandeker04}.

In this paper we present the first detection of atomic Hydrogen in the gas disk surrounding $\beta$\,Pic. In Sect.\,\ref{sec:observations} we present the observations and describe our new technique for reducing the airglow contamination. The analysis of the data is described in Sect.\,\ref{sec:analysis} with a description of how the emission line profile was modelled and how we calculated the column density of hydrogen. The results and discussion are presented in Sect.\,\ref{sec:results} and Sect.\,\ref{sec:discussion} respectively, with a discussion on the possible origin of the hydrogen gas. A final conclusion summarising the main findings is presented in Sect.\,\ref{sec:conclusions}.

\section{Observations}
\label{sec:observations}
Far-UV observations of $\beta$\,Pic were obtained using the {\it{Cosmic Origins Spectrograph (COS)}} on the {\it{Hubble Space Telescope (HST)}} using the TIME-TAG mode and the G130M grating. The observations were done using the primary science aperture which has a 2.5\,\arcsec diameter field stop. The observations consisted of a total of four visits, each divided into two {\it{HST}} orbits where the first orbit was aimed at observing the Ly-$\alpha$ line of $\beta$\,Pic and the second orbit aimed at measuring the airglow by observing the sky emission only, one arcminute away from $\beta$\,Pic. The visits were conducted on 24 February 2014, 10 December and 26 December 2015 and on 30 January 2016.

\subsection{Contamination by airglow}
Terrestrial hydrogen, oxygen and nitrogen in Earth's exosphere produce emission lines, known as airglow, which become superimposed on the $\beta$\,Pic spectrum. Before any scientific exploitation of the airglow contaminated data can begin the Ly-$\alpha$ airglow emission has to be removed. From previous Ly-$\alpha$ COS observations of the exoplanet host star HD189733 (\citealt{BenJaffel13} and HST observations \#11673, PI Lecavelier) it was demonstrated that the airglow contamination is very stable from one HST orbit to the other, producing variable but reproducible features as a function of HST orbital position. We performed airglow measurements during the second orbit of each visit to determine a map of the airglow contamination as a function of wavelength and remove it from the data.

   \begin{figure}
   \centering
   \includegraphics[width=\hsize]{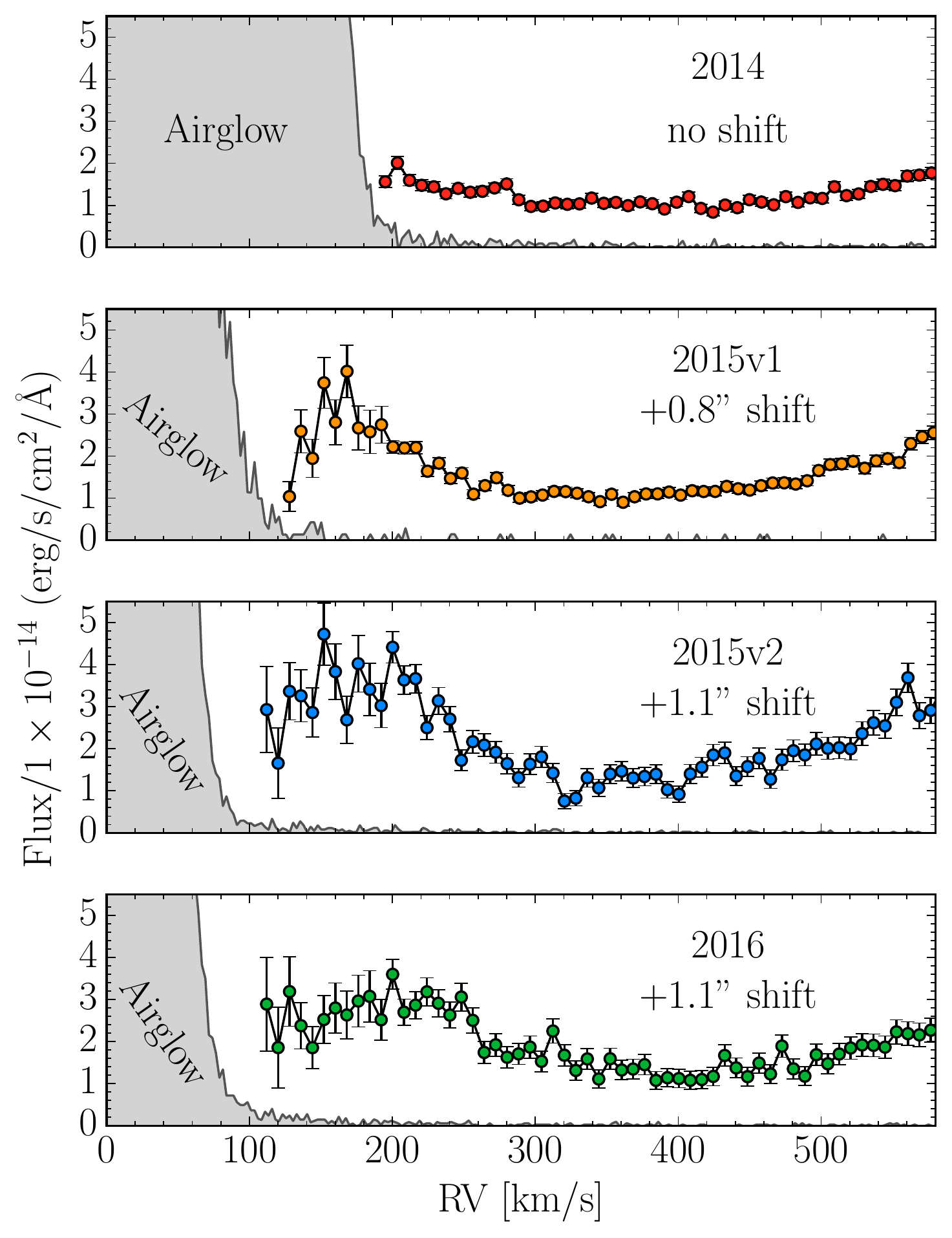}
      \caption{The red wing of the $\beta$\,Pic Lyman-$\alpha$ profile showing data not affected by airglow contamination. The measured flux is shown as a function of RV with respect to the $\beta$\,Pic reference frame. For sake of clarity each point has been created though calculating the average of three individual measurements. The "hook"-like feature seen for radial velocities less than $\sim150$\,km/s is due to the absorption by the hydrogen present in the CS gas surrounding $\beta$\,Pic. The datapoints covering the line core are not shown due to airglow contamination. Increasing the offset along the dispersion axis virtually move the airglow away and reveals more of the Ly-$\alpha$ profile. No shift was applied in 2014 resulting in the red points in the top panel which show only the start of the profile before the data is contaminated by the airglow. An 0.8\,\arcsec shift along the dispersion axis reveals the Ly-$\alpha$ down to $\sim122$\,km/s as shown by the data obtained during the first visit in 2015 (orange points shown in second panel from the top). For the largest shift of 1.1\,\arcsec done during the second visit of 2015 and the visit of 2016 (blue and green points in the lower panels respectively) the profile is revealed down to 115\,km/s. 
              }
         \label{fig:Ly_cut}
   \end{figure}

\subsection{A new airglow contamination reduction technique: Airglow Virtual Motion}
For the observations done on December 10 2015 we introduced a new technique, called the Airglow Virtual Motion (AVM) technique, designed to limit the airglow contamination of the $\beta$\,Pic spectrum. With this new technique being successful, we subsequently used it for the remaining observations done on December 26 2015 and 30 January 2016. The AVM technique consists of an off-axis placement of the target (here $\beta$\,Pic) away from the central position of the aperture in the dispersion direction, in order to shift the target spectrum in wavelength and thus partially separate the target spectrum from the airglow spectrum. As the target is placed closer to the edge of the circular aperture we effectively alter the location of the target spectrum on the detector. The airglow, however, which fills the entire aperture regardless of any induced shift, remains constant on the detector and is thus virtually moved away from the target Ly-$\alpha$ spectrum. This comes at a cost of a decreased flux as less of the light from the point source target passes through the aperture. To limit the airglow contamination of the Ly-$\alpha$ line wings, the observations were conducted with $\beta$\,Pic placed -0.8, +0.8 and +1.1 arcseconds away from the central position of the aperture (-0.8 and +0.8 arcseconds for the first visit of 2015 and -0.8, +0.8 and +1.1 arcseconds for the last visit of 2015 and the visit of 2016). A shift in the positive dispersion direction shifts the target spectrum redward and a shift in the negative dispersion direction shifts the target spectrum blueward. The AVM technique successfully revealed more of the non-contaminated $\beta$\,Pic Ly-$\alpha$ line profile with a greater shift allowing more data towards the core to be uncovered (see Fig.\ref{fig:Ly_cut}). For a more technical description of this method see Appendix \ref{sec:appendix}.

\section{Analysis}
\label{sec:analysis}
\subsection{Wavelength calibration of data}
The data was reduced using the CALCOS pipeline \citep{fox15} version 2.20.1 for the 2014 data and version 3.1.1 for the 2015 and 2016 data. The offset of $\beta$\,Pic along the dispersion axis, executed on purpose to reveal more of the stellar Ly-$\alpha$ emission, caused the $\beta$\,Pic spectra to be shifted. The more $\beta$\,Pic was offset in the aperture, the larger the spectral shift was. To wavelength calibrate the shifted $\beta$\,Pic spectra we calculated their shift by cross correlating the shifted spectra with the non-shifted 2014 $\beta$\,Pic spectrum. The section of the spectra used in the cross correlation calculation were carefully selected based on three criteria: the absence of airglow contamination, the absence of exocomet activity and a relatively high signal to noise ratio. Once the spectra had all been shifted to the 2014 spectral reference we divided the spectrum obtained in 2014 by the newly aligned spectra obtained in 2015 and 2016 to look for large deviations from unity. This was done to more easily detect weak exocomet signatures. We used the divided spectra to again adjust the best region before repeating the cross-correlation and shifting process once more. The region from 1224.4 to 1259.0\,\AA\ was found to be the optimal region for cross-correlation with a signal-to-noise favorable for an accurate shift to be calculated. The off-axis placement of $\beta$\,Pic away from the centre of the aperture resulted in a flux loss; with the $\pm$0.8 arcsecond and a +1.1 arcsecond shifts resulting in a flux decrease of 22 and 43\,\% respectively (see Appendix \ref{sec:appendix}). We calculated the difference in flux between the 2014 observation and the remaining spectra by calculating the median flux value for all of the spectra across a wavelength range close to the Ly-$\alpha$ emission line (1208.45 to 1213.43\,\AA) known to be void of strong spectral lines. Using the unshifted 2014 data as a reference we multiplied the remaining spectra by the ratio of the medians between the 2014 data and each individual spectrum individually to bring them all to the same flux level. Redward of the stellar Ly-$\alpha$ emission we detect variations in the stellar O\,\Rmnum{5} line at 1218.3440\,\AA\ (see Fig.\,\ref{fig:Ly_cut}) . The variability of the O\,\Rmnum{5} does not affect the absorption of the H\,\Rmnum{1} line. This is because the absorption depth and shape is calculated relative to the stellar profile. An increase in emitted photons causes an increase of the number of absorbed photons.

   \begin{figure}
   \centering
   \includegraphics[width=\hsize]{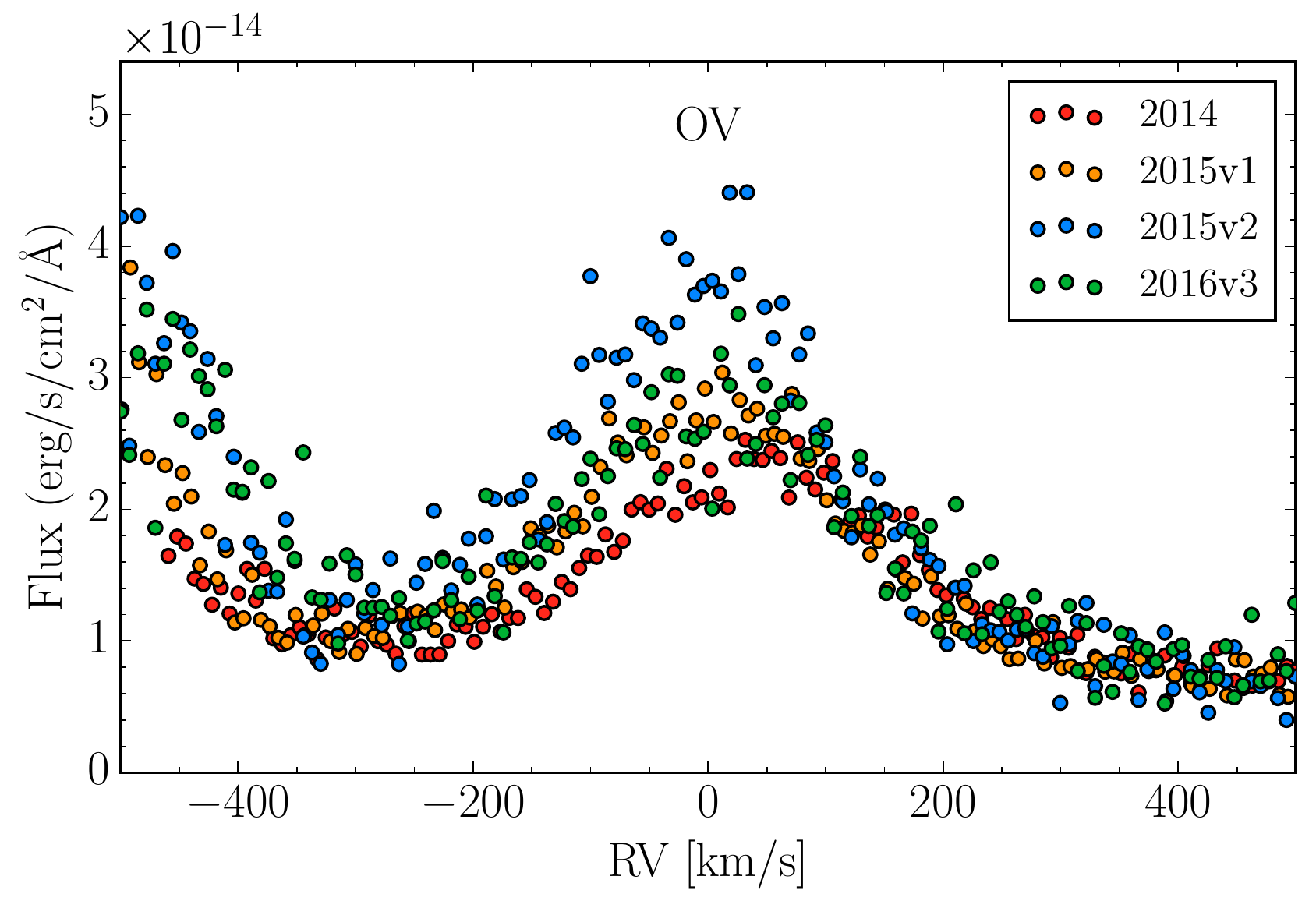}
      \caption{The stellar O\,\Rmnum{5} line at 1218.3440\,\AA\ line shown as a function of RV with respect to the O\,\Rmnum{5} reference frame. The variations in amplitude could be due to stellar activity.
              }
         \label{fig:Ly_cut}
   \end{figure}

\subsection{Uncertainties on flux measurements}
The uncertainties produced by the CALCOS pipeline were found to be overestimated by a factor of several (depending on the visit). The overestimate was largest for measured flux levels below approximately $10^{-14}$\,erg/s/cm$^2$/\AA, when the number of detected photons per spectral bin amounts to only a few. % or zero for some pixels.
This was noticed when comparing the uncertainties with the dispersion of the measurements. The dispersion was found to be consistent with a noise budget dominated by photon noise. We calculated the uncertainties using the following equation,

\begin{equation}
\label{eq:main}
\mathrm{Err} = \sqrt{(C\times T+1)} \times S%\frac{F\times C}{E(C-B)\times T}
\end{equation}

\noindent where $C$ is the gross count rate in [counts s$^{-1}$] (counts from all sources including background counts), $T$ the exposure time in [s] and $S$ the sensitivity of the detector in [erg s$^{-1}$ cm$^{-2}$ \AA$^{-1}$counts$^{-1}$] at a given pixel. A tabulated $S$ is used in the pipeline reduction process. We retrieve this value from the resulting output of the pipeline using the equation

\begin{equation}
\label{eq:main}
S = \frac{F}{net\times T}
\end{equation}

\noindent where $F$ is the tabulated flux in [erg s$^{-1}$ cm$^{-2}$ \AA$^{-1}$] and $net$ is the difference between $C$ and the background count rate with flat-field and dead-time effects taken into account in [counts s$^{-1}$]. To be conservative in our error estimate we added "$+1$" to the product $C\times T$ to avoid the uncertainty becoming zero when the number of photons become low. This does not occur in the wavelength region  considered here.

   \begin{figure}
   \centering
   \includegraphics[width=\hsize]{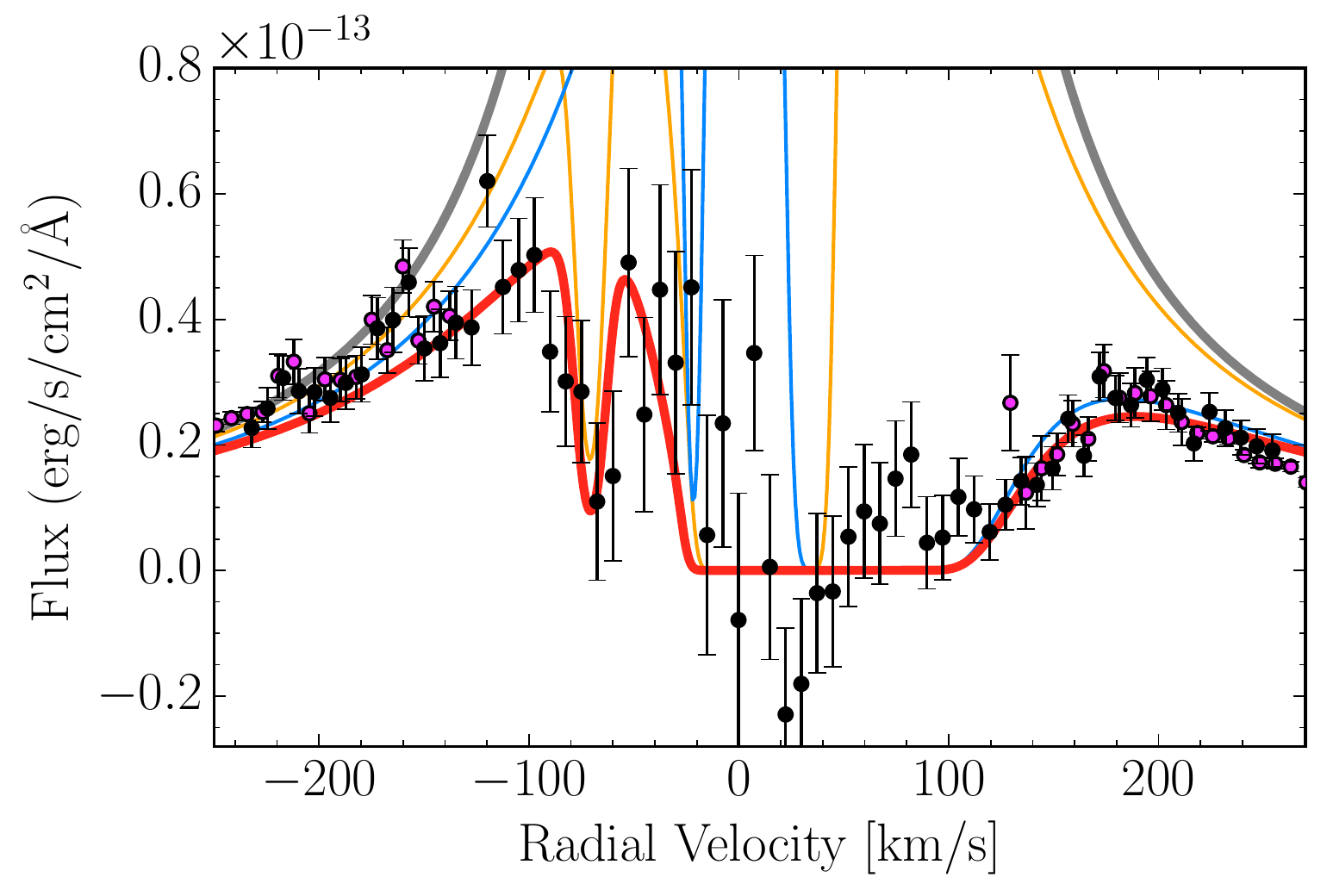}
      \caption{Measured flux of the Ly-$\alpha$ emission line as a function of radial velocity in the $\beta$\,Pic reference frame. For sake of clarity each data point has been created by calculating the average of three individual measurements. The data obtained using the AVM technique and ignoring the airglow contaminated regions ({\it{airglow-free}}) is shown as pink points. The sky subtracted ({\it{airglow-corrected}}) data is shown as black points which represent the weighted mean of each individual visit. The best fit model based on the {\it{airglow-corrected}} data is shown as a red line. This profile was calculated by subtracting the ISM absorption (ISM-absorbed line shown in yellow) and the absorption by the CS gas (CS-gas-absorbed line shown in blue) away from the stellar Lyman-$\alpha$ emission line (thick grey line). The final profile (red line) is convolved with the instrumental line spread function taken to be Gaussian with a full width at half maximum (FWHM) of 6.5 pix.
              }
         \label{fig:Ly_alpha}
   \end{figure}

\subsection{Airglow correction}
\label{airglow_correct}

\subsubsection{Method}
We removed the contamination by airglow using the following method:
\begin{enumerate}
\item Create an airglow map with the weighted average of all the airglow spectra, $\overline{AG}_w$, obtained during the second orbit of each visit.
\item Calculate a scaling factor $f$, which determines how much of $\overline{AG}_w$ to subtract from the airglow contaminated spectra, $\beta_{\mathrm{AG}}$, in order to obtain the airglow free spectrum $\beta_{\mathrm{F}}$, i.e. $\beta_{\mathrm{F}} = \beta_{\mathrm{AG}} - f\times \overline{AG}_w$. The scaling factor is found by calculating the median of the division of $\beta_{\mathrm{AG}}$ by $\overline{AG}_w$ as a function of radial velocity over the region where the stellar Ly-$\alpha$ line is completely absorbed by the ISM and $\beta_{\mathrm{F}}$ is known to be zero.
\item Finally subtract the airglow contamination using $\beta_{\mathrm{F}} = \beta_{\mathrm{AG}} - f\times \overline{AG}_w$.
\end{enumerate}

\noindent For the first step all airglow data was used except the airglow spectrum obtained on the 26 December 2015 which displayed a sharp discontinuity probably due to systematics. For the second step we calculated $f$ assuming the flux is zero within the radial velocity range from 0.5 to 40.5 km/s. This range was conservatively selected based on the data unaffected by the airglow. The best fit using the airglow corrected data shows that the region of zero flux is likely larger and the used region is therefore conservative. We assessed the impact of choosing a different radial velocity range over which we hypothesise the flux remains zero and found that the calculated CS column density remains constant providing the velocity range does not exceed 160\,km/s. Beyond a range of 160\,km/s results in a warped profile which produces a worse fit. The uncertainty was calculated as $E_{\beta_{\mathrm{F}}} = \sqrt{E_{\mathrm{AG}}^2+\left (f\times E_{\overline{AG}_w}\right )^2}$ where $E_{\mathrm{AG}}$ and $E_{\overline{AG}_w}$ are the per-pixel uncertainties on $\beta_{\mathrm{AG}}$ and $\overline{AG}_w$ respectively.

\begin{table}%[h]
 \caption[]{Parameters derived from fitting the Ly-$\alpha$ line.}
\label{table:params}
\begin{tabular}{lcc}
 \hline \hline
\noalign{\smallskip}
Parameter & Value & Free/Fixed\\
\hline
\noalign{\smallskip}
\noalign{\smallskip}
\multicolumn{3}{c}{Stellar Ly-$\alpha$ line parameters (emission)} \\
\noalign{\smallskip}
\hline
\noalign{\smallskip}
$E_{\mathrm{max}}$                      &   $4.2^{+0.5}_{-0.4}\times10^{-10}$\,erg/s/cm$^2$/\AA  & free \\
\noalign{\smallskip}
$\sigma_{\mathrm{Ly}\alpha}$     &   $\sim0.003$\,\AA\ / $0.7$\,km/s     & free  \\
\noalign{\smallskip}
$\gamma_{\mathrm{Ly}\alpha}$    &   $\sim0.022$\,\AA\ / $5.5$\,km/s   & free  \\
\noalign{\smallskip}
$v_{\beta\,\mathrm{Pic}}$                 &   $20.5$\,km/s   & fixed \\
\noalign{\smallskip}
\hline

\noalign{\smallskip}
\noalign{\smallskip}
\multicolumn{3}{c}{ISM (absorption)} \\
\noalign{\smallskip}
\hline
\noalign{\smallskip}
$\log(N_{\mathrm{H}}/1\,\mathrm{cm}^2)_{\mathrm{ISM}}$   &    $18.2\pm0.1$\,cm$^{-2}$&   free  \\
\noalign{\smallskip}
$T_{\mathrm{ISM}}$                      &   7000\,K                 &   fixed   \\
\noalign{\smallskip}
$\xi_{\mathrm{ISM}}$                      &   1.5\,km/s                 &   fixed   \\
\noalign{\smallskip}
$v_{\mathrm{ISM}}$                      &   $10\pm5$\,km/s                &   fixed   \\
\noalign{\smallskip}
\hline

\noalign{\smallskip}
\noalign{\smallskip}
\multicolumn{3}{c}{CS Gas (absorption)} \\
\noalign{\smallskip}
\hline
\noalign{\smallskip}
$\log(N_{\mathrm{H}}/1\,\mathrm{cm}^2)_{\mathrm{CS}}$  &   $18.6\pm0.1$ &   free    \\
\noalign{\smallskip}
$v_{\mathrm{CS}}$                     &   $41\pm6$\,km/s (relative to $\beta$ Pic)  &   free    \\
\noalign{\smallskip}
$\xi_{\mathrm{CS}}$                     &   $6^{+4}_{-3}$\,K                 &   free   \\
\noalign{\smallskip}
$T_{\mathrm{CS}}$                     &   $3700^{+4900}_{-2700}$\,K                 &   free   \\
\noalign{\smallskip}
$b_{\mathrm{CS}}$                     &   $8^{+4}_{-3}$\,km/s                 &   calculated\tablefootnote{The total broadening parameter, $b$, is calculated using $b^2=2kT/m+\xi$$^2$ where $k$ is the Boltzmann constant, $m$ the mass of the considered species and $\xi$ the turbulent velocity parameter (the non-thermal broadening parameter).}    \\
\noalign{\smallskip}
\hline
\noalign{\smallskip}

%$^\dagger$ $b^2$=2kT/m+$\xi$$^2$ where $k$ is the Boltzmann constant and $m$, the mass\\
%of the considered species and $\xi$ the turbulent velocity\\
%paramter (non-thermal broadening parameter)

\end{tabular}

\end{table}

\subsubsection{Validating the airglow correction.}
To check the validity of the airglow removal technique we constructed a Ly-$\alpha$ profile only using data which was not contaminated by airglow and ignoring data in the core of the line which was. We compare this {\it{airglow-free}} profile to the {\it{airglow-corrected}} profile which was created by subtracting the airglow contamination from the $\beta$\,Pic spectra following the steps presented in Sect.,\ref{airglow_correct}. In the {\it{airglow-free}} case we ignored any data contaminated by airglow (resulting in the pink points in Fig.\,\ref{fig:Ly_alpha}). In the {\it{airglow-corrected}} case we use the dedicated airglow measurements to subtract the airglow contamination from the data and do not discard any data (resulting in the black points in Fig.\,\ref{fig:Ly_alpha}). To create a profile unaffected by airglow, the {\it{airglow-free}} profile, we recorded the shift induced on each of the $\beta$\,Pic spectra which resulted from off-axis observations. Each of the airglow spectra obtained during the second orbit of each visit were then purposely offset by the same recorded shift noted for each of the $\beta$\,Pic spectra. By comparing the $\beta$\,Pic spectra (now with the correct wavelength) against the airglow spectrum (shifted by the same amount as the $\beta$\,Pic spectra previously was) we were able to determine which part of the $\beta$\,Pic spectrum was affected by the airglow. We find that the contamination by airglow starts to become noticeable when the measured airglow flux exceeds half the measured uncertainty of $\beta$\,Pic spectra ($F_{\mathrm{AG}} > 0.5\,\beta_{\mathrm{err}}$). Data points within the region affected by Ly-$\alpha$ airglow were removed before the data was combined using a weighted mean. Fig.\,\ref{fig:Ly_cut} shows the $\beta$\,Pic spectra of each individual visit. It is clear that for the observations done in 2015 and 2016, the technique of placing the target off-centre in the aperture (AVM technique) allowed us to resolve more of the Ly-$\alpha$ emission by virtually moving the region contaminated by airglow. The scatter of the data points increase towards the central part of the line as fewer observations were done with the 0.8 and 1.1\,\arcsec offsets. The "hook" like feature at $\sim150$\,km/s, observed outside the airglow contaminated region during three separate visits, indicates a clear detection of hydrogen absorption in the centre of the stellar Ly-$\alpha$ emission line. The 2014 observations had no offset applied and thus could not probe the line below $\sim190$\,km/s, thereby missing the "hook" like feature. The combined data from the individual visits are shown as pink points in Fig.\,\ref{fig:Ly_alpha}.

The two profiles, {\it{airglow-free}} and {\it{airglow-corrected}} produce consistent results as is shown in Fig.\,\ref{fig:Ly_alpha}. This is hardly surprising as much of the same data is used to construct both profiles. It does, however, demonstrate that only the inner region of the Ly-$\alpha$ is affected by airglow. It also shows that the detection of hydrogen as an absorber below $\sim150$\,km/s is robust, as indicated by the hook like feature present in all the data (except for the data from 2014 which was not sensitive to below $\sim190$\,km/s). Since the hydrogen absorption is seen in both the {\it{airglow-free}} and {\it{airglow-corrected}} profiles we conclude that the detected presence of hydrogen in the $\beta$\,Pic environment is not dependent on how we treat the airglow contamination.

   \begin{figure*}
   \centering
   \includegraphics[width=\hsize]{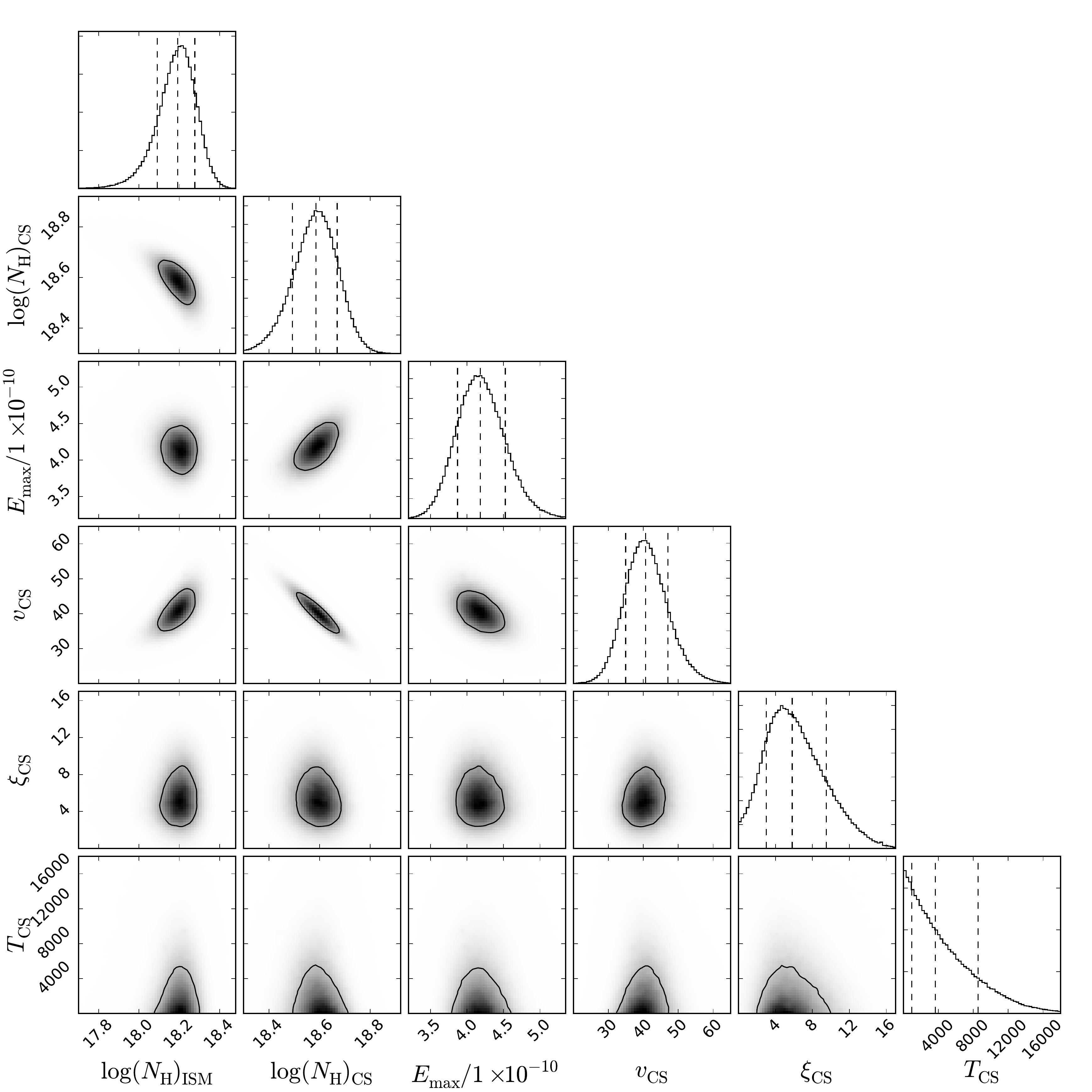}
      \caption{A corner plot showing the one and two dimensional projections of the posterior probability distributions for the free parameters. The dashed lines in each of the 1D histograms represent the 1\,$\sigma$ deviations (68\,\% of the mass) with the central dashed line indicating the median value. The solid black lines in each of the 2D histograms represent the 1\,$\sigma$ level (39.3\,\% of the volume).
              }
         \label{fig:mcmc}
   \end{figure*}

\subsection{Fit to the data and error bars}

The best fit and the associated uncertainties of the individual parameters were calculated using the Markov chain Monte Carlo (MCMC) method. The merit function was calculated using the $\chi^2$ of the difference between the data and the model in the wavelength range from 1214.7\,\AA\ to 1216.7\,\AA , corresponding the range from -260\,km/s to 235\,km/s in radial velocity in the $\beta$\,Pic rest frame. The best $\chi^2$ is found to be 197 for 196 degrees of freedom. 

We ran 24 MCMC chains with a total of $1\times10^6$ accepted steps, with an average acceptance rate of $\sim25$\,\%. The posterior probability distributions for the free model parameters are shown in Fig.\,\ref{fig:mcmc}, along with the marginalised 1D distributions which were used to calculate the median values and the uncertainties on each free parameter (see Table\,\ref{table:params}). In the next section we describe the model used and our choice of free model parameters.

\section{Results}
\label{sec:results}

\subsection{Modeling the $\beta$\,Pic Ly-$\alpha$ line profile:}
\label{sec:model}

In the construction of a $\beta$\,Pic Ly-$\alpha$ line profile we considered a model consisting of a stellar Ly-$\alpha$ emission line profile absorbed by two separate components; the ISM and the CS gas. For the second of these two absorbing components, the CS gas, we investigated two possible options. The first was a single absorbing CS gas component with a bulk velocity set to vary freely. The second were two absorbing CS gas components, one fixed at the at the radial velocity of $\beta$\,Pic with an additional component with a bulk velocity parameter free to vary. To calculate an analytic line profile for the $\beta$\,Pic Ly-$\alpha$ line we used the following equation

\begin{multline}
\label{eq:model}
\mathrm{F}(\lambda) = \left ( \mathrm{C}(\lambda) + \mathrm{E}_{\beta \mathrm{\,Pic}}(\lambda)\right ) \times \mathrm{A}_{\mathrm{ISM}}(\lambda,N_\mathrm{H, ISM},\xi,T) \\
 \times \mathrm{A}_{\mathrm{CS}}(\lambda,N_\mathrm{H, disk},\xi,T)
\end{multline}

\noindent where,

\begin{equation}
\label{eq:C}
\mathrm{C}(\lambda) = \sum_i^N \mathrm{C}_i \lambda^{i}
\end{equation}

\noindent and

\begin{equation}
\label{eq:F}
\mathrm{E}_{\beta \mathrm{\,Pic}}(\lambda) = E_{\mathrm{max}}\times\mathrm{Voigt}(\lambda,\gamma,\sigma).
\end{equation}

\begin{figure*}
   \centering
   \includegraphics[width=\hsize]{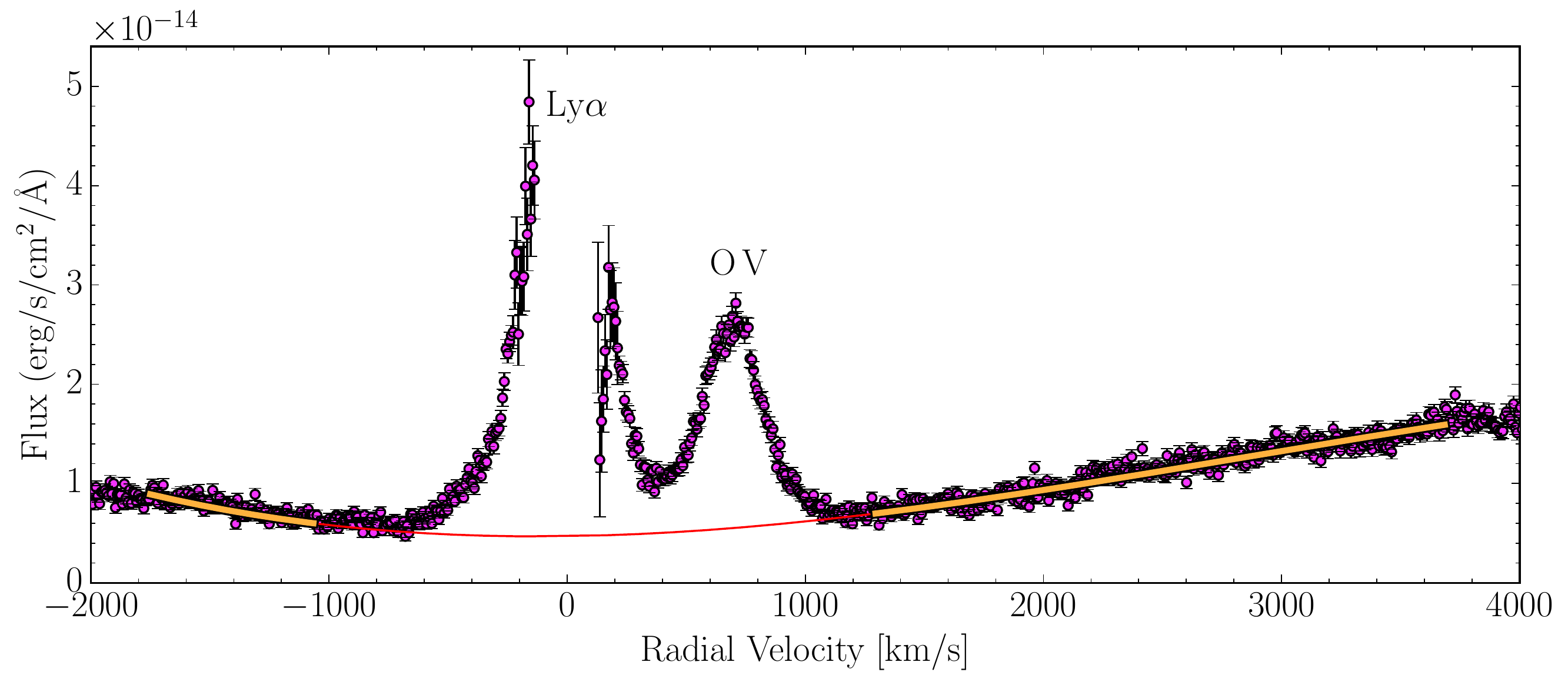}
      \caption{Measured flux as a function of radial velocity with respect to the Ly-$\alpha$ emission line in the $\beta$\,Pic reference frame. For sake of clarity each data point (pink points) has been created by calculating the average of three individual measurements. The data contaminated by airglow has been ignored ({\it{airglow-free}}) leaving a gap in the central part of the Ly-$\alpha$ profile. To model the very broad stellar Ly-$\alpha$ absorption line profile (see Sect.\,\ref{sec:model}) we fit the flux continuum using a 4$^{\mathrm{th}}$ order polynomial to produce a continuum model (thin red line) using the wavelength range covered by the bold orange line. Above the broad stellar absorption line, we see the narrower Ly-$\alpha$ emission line. The feature to the right of the Ly-$\alpha$ profile is the O\,\Rmnum{5} line at 1218.3440\,\AA\ ($\sim700$\,km/s).
              }
         \label{fig:AG_free}
   \end{figure*}

\noindent The observed flux, $\mathrm{F}(\lambda)$, consists of the addition of a broad stellar absorption line, $\mathrm{C}(\lambda)$ (seen in Fig.\,\ref{fig:AG_free}), and a narrower stellar Ly-$\alpha$ emission line, $\mathrm{E}_{\beta \mathrm{\,Pic}}$ (seen in Fig.\,\ref{fig:Ly_alpha}), which is absorbed by the deuterium and hydrogen gas in the ISM ($\mathrm{A}_{\mathrm{ISM}}$) and the $\beta$\,Pic environment ($\mathrm{A}_{\mathrm{CS}}$). The Ly-$\alpha$ emission line was calculated with the peak flux, $E_{\mathrm{max}}$, standard deviation of the Gaussian part of the Voigt profile, $\sigma_{\mathrm{Ly}\alpha}$, and the half-width at half-maximum of the Lorentzian part of the Voigt profile, $\gamma_{\mathrm{Ly}\alpha}$, as free parameters.

The parameters used to model the ISM absorption; $T_{\mathrm{ISM}}$ (the temperature of the ISM), $\xi_{\mathrm{ISM}}$ (the turbulent velocity of the ISM) and $v_{\mathrm{ISM}}$ (the bulk velocity of the gas), were all kept fixed except for $\log(N_{\mathrm{H}}/1\,\mathrm{cm}^2)_{\mathrm{ISM}}$ (the ISM column density) which was set as a free parameter.

To obtain the $\beta$\,Pic gas parameters the two options for modeling the $\beta$\,Pic CS gas were explored. The first option, consisted of a single absorbing CS gas component with $\log(N_{\mathrm{H}})_{\mathrm{CS}}$, $v_{\mathrm{CS}}$, $\xi_{\mathrm{CS}}$ and $T_{\mathrm{CS}}$ set as free parameters. The bulk velocity of the CS gas component, $v_{\mathrm{CS}}$, was chosen as a free parameter as this produced a significantly better fit compared to having it fixed at the radial velocity of $\beta$\,Pic. The temperature, $T_{\mathrm{CS}}$, and the turbulent velocity $\xi_{\mathrm{CS}}$, do not significantly impact the determination of $\log(N_{\mathrm{H}})_{\mathrm{CS}}$. This is because the H\,\Rmnum{1} line is heavily saturated and the shape of the absorption profile is completely dominated by the damping wings. The second option, consisting of two $\beta$\,Pic gas components, was chosen to be identical to the first option except that one of the components had $v_{\mathrm{\beta}}$ fixed at the same radial velocity as $\beta$\,Pic with an additional component $v_{\mathrm{X}}$ set to vary freely. This second option resulted in a CS column density of the first component fixed at 20.5\,km/s to be $\log(N_{\mathrm{H}}/1\,\mathrm{cm}^2)_{\beta} = 17.2$ and the second component with a freely varying bulk velocity to be $\log(N_{\mathrm{H}}/1\,\mathrm{cm}^2)_{X}=18.6$. To estimate the total column density of H gas we add the components

\begin{equation}
\label{eq:col_add}
\begin{split}
\log(N_{\mathrm{H}}/1\,\mathrm{cm}^2)_{\mathrm{CS}} &= \log(10^{N_{\mathrm{H}\beta}/1\,\mathrm{cm}^2} + 10^{N_{\mathrm{H}X}/1\,\mathrm{cm}^2})\\
&= \log(10^{17.2}+10^{18.6}) = 18.6
\end{split}
\end{equation}

\noindent and obtain the same column density as the first option. This demonstrates that the same hydrogen column density value is obtained regardless of the model. Due to a strong degeneracy between the two components in the second model option caused by the line center saturation, we decided to use the simpler option of the one component model. We note that there are many other possible solutions given by other models consisting of more components. However, we can not justify the added complexity and therefore choose the simpler model. The first model option has a lower BIC (Bayesian information criterion) value, 227 compared to 237 for the second option. The parameters used for the fit are all shown in Table\,\ref{table:params} with the posterior distribution of the most relevant parameters in Fig.\,\ref{fig:mcmc}.

It is likely that the Ly-$\alpha$ lines is subject to a certain amount of variability. Despite this, the variability is not expected to have any impact on the derived HI abundance as the absorption depth and shape is calculated relative to the stellar profile.

\subsection{Interstellar medium}
\label{sec:Interstellar medium}

The interstellar medium toward $\beta$\,Pic is detected in several lines such as Fe\,\Rmnum{2}, Mg\,\Rmnum{1} and Ca\,\Rmnum{2} \citep{vidal-madjar_1994, lallement95}. Its radial velocity corresponds to the radial velocity of the Local Interstellar Cloud (LIC) that is at +10\,km/s in the heliocentric reference frame \citep{lallement95}, which corresponds to -10\,km/s relative to $\beta$\,Pic. A temperature of 7000\,K and $\xi_{\mathrm{ISM}}=1.5$\,km/s was chosen based on measurements of the LIC cloud towards Capella and Procyon \citep{bertin95, linksy95} and measurements of Fe\,\Rmnum{2} \citep{lallement95} in the line of sight towards $\beta$\,Pic; none of the results discussed here depends on this assumption.

Importantly, a narrow absorption feature appears at $\sim-90$\,km/s in the $\beta$\,Pic reference frame
(see Fig.\,\ref{fig:Ly_alpha}). This feature corresponds exactly to the expected absorption by the interstellar deuterium D\,\Rmnum{1} line. The presence of this line strengthens the confidence in the airglow correction methods described above. Moreover, this shows that the column densities of both ISM and the $\beta$\,Pic circumstellar gas must be low enough ($N\la 3\times 10^{19}$\,cm$^{-2}$) to avoid wide damping wings in their absorption profiles, which would have hidden this narrow deuterium line. 

We modelled the narrow D\,\Rmnum{1} line independently of the H\,\Rmnum{1} line to determine its origin. The best fit was achieved with a radial velocity of $10\pm5$\,km/s. We conclude that the detected absorption is due to D\,\Rmnum{1} in the ISM and not related to the $\beta$\,Pic circumstellar gas disk at the 2\,$\sigma$ confidence level. A simultaneous fit of the D\,\Rmnum{1} and H\,\Rmnum{1} lines, assuming D/H ratio of 1.5$\times$$10^{-5}$ \citep{hebrard02, linsky06}, yields a H\,\Rmnum{1} ISM column density of $\log(N_{\mathrm{H}}/1\,\mathrm{cm}^2)_{\mathrm{ISM}} = 18.2\pm 0.1$. This column density provides a good fit of both the narrow D\,\Rmnum{1} line and the blue side of the broad absorption feature (Fig.\,\ref{fig:Ly_alpha}).

%D/H ratio of 1.5$\times$$10^{-5}$ \citep{hebrard02, linsky06}, {\bf{we do a simultaneous fit of the D\,\Rmnum{1} and H\,\Rmnum{1} lines and find that}} the best fit is obtained for an H\,\Rmnum{1} ISM column density of $\log(N_{\mathrm{H}}/1\,\mathrm{cm}^2)_{\mathrm{ISM}} = 18.2\pm 0.1$.

This result is in total agreement with previous estimates of ISM column density in the solar neighborhood. 
For instance, from the compilation by \cite{linsky06} of column density measurements of interstellar gas along the lines of sights of 47 stars we find that targets at a distance $d \leq 19\,\mathrm{pc}$ \cite{vanLeeuwen07} all have $\log(N_{\mathrm{H}}/1\,\mathrm{cm}^2)_{\mathrm{ISM}}< 18.3$. 
Our measurement of $\log(N_{\mathrm{H}}/1\,\mathrm{cm}^2)_{\mathrm{ISM}} = 18.2\pm 0.1$ is also in good agreement with work by \cite{wood05} who measured the ISM H\,\Rmnum{1} column densities of 33 Ly-$\alpha$ spectra from the {\it{HST}} archive.

%Finally, {\bf{to test that the narrow D\,\Rmnum{1} line has an interstellar origin, we model the line profile independently of the H\,\Rmnum{1} line. We found the heliocentric radial velocity of the corresponding absorption to be $10\pm5$\,km/s.}} This excludes the possibility that the detected absorption is due to the $\beta$\,Pic circumstellar gas disk at 2\,$\sigma$ confidence level. 

%by modeling the narrow D\,\Rmnum{1} line, we are able to constrain the heliocentric radial velocity of the corresponding absorption to be $10\pm5$\,km/s.

\subsection{Asymmetry of the Ly-$\alpha$ profile}
\label{sec:Asymmetry of the Ly-alpha profile}

The detected Ly-$\alpha$ emission profile shows a significant asymmetry, with a peak flux on the blue side ($\sim 4\times 10^{-14} $\,erg/s/cm$^2$/\AA) almost twice the flux on the red side ($\sim 2.5\times 10^{-14} $\,erg/s/cm$^2$/\AA) (see Fig.\,\ref{fig:Ly_alpha}). To explain this asymmetry, either a blueshifted emission in addition to the stellar chromospheric line or a redshifted absorption over this stellar line is required. This cannot be an absorption due to the ISM, which is known to be blueshifted by -10\,km/s relative to $\beta$\,Pic \citep{lallement95}, and we cannot identify a known emission process able to produce a blueshifted emission. The asymmetric shape of the observed Lyman-$\alpha$ line can only be explained by the presence 
of hydrogen gas in the line of sight producing redshifted absorption by moving in towards the star.

\subsection{Redshifted absorbing hydrogen}

   \begin{figure}
   \centering
   \includegraphics[width=\hsize]{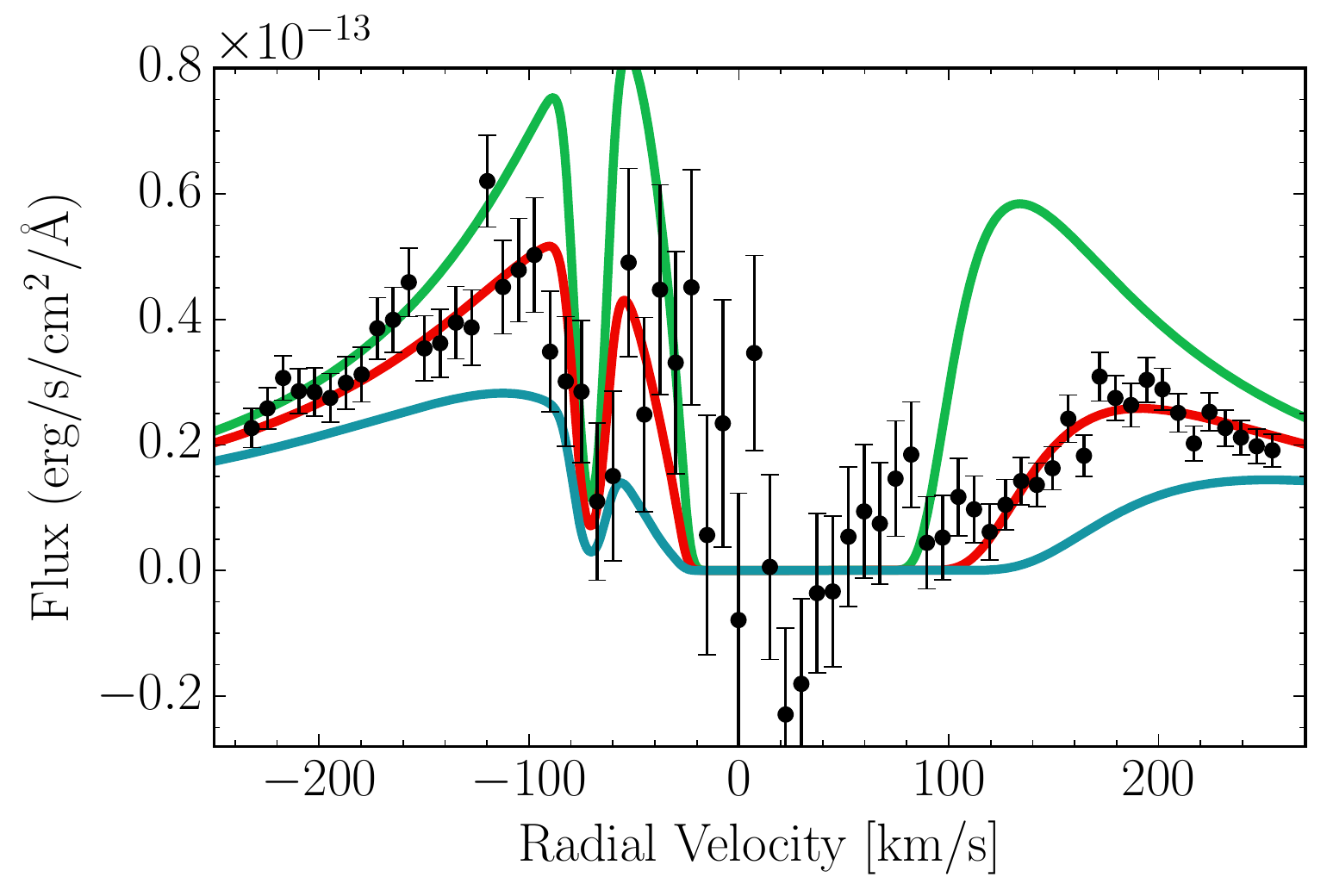}
      \caption{Measured flux of the Ly-$\alpha$ emission line as a function of radial velocity in the $\beta$\,Pic reference frame. For sake of clarity each data point has been created by calculating the average of three individual measurements. The lines in green, red and blue correspond to values of $\log(N_{\mathrm{H}}/1\,\mathrm{cm}^2)$ of the $\beta$\,Pic absorption component equal to 18.0, 18.6 and 19.0 respectively, whilst keeping all other parameters constant. Fits generated with $\log(N_{\mathrm{H}}/1\,\mathrm{cm}^2)$ different from the best fit at 18.6 produce significantly worse fits.
              }
         \label{fig:vary_param}
   \end{figure}

The presence of redshifted absorbing gas has been observed toward $\beta$\,Pic over the course of three decades 
and is explained by the presence of {\it Falling Evaporating Bodies}, or FEBs, in other words 
``exocomets'' \cite{ferlet_1987, beust_1990, kiefer14}. 
Nonetheless, these exocomets produce variable features, while we detect no significant 
variations in the H\,\Rmnum{1} profiles observed at four epochs spanning almost two years.
In short, we detect atomic hydrogen moving in towards the star; it remains unknown if this gas is produced by
a family of exocomets passing in front of the star on similar orbits to the D-family of exocomets identified by \cite{kiefer14} in the Ca\,\Rmnum{2} lines, but this appears to be a possibility.

Setting the bulk velocity of the hydrogen absorption free to vary, we obtained a best fit with a radial 
velocity of $41\pm6$\,km/s relative to the stellar reference frame. For this single component, we calculate a hydrogen column density of $\log(N_{\mathrm{H}}/1\,\mathrm{cm}^2) = 18.6\pm0.1$. This value is compatible with the upper limit set by radio observations of the 21\,cm emission line by \cite{freudling95}. Adjusting the value of $\log(N_{\mathrm{H}}/1\,\mathrm{cm}^2)$ even slightly causes a significantly worse fit as is demonstrated in Fig.\,\ref{fig:vary_param}.

\subsection{Circumstellar gas disk at zero radial velocity}
\label{sec:Circumstellar gas disk at zero radial velocity}

Beyond the need for a single redshifted absorption component to explain the Lyman-$\alpha$ profile asymmetry
(see Sect.~\ref{sec:Asymmetry of the Ly-alpha profile}), 
we tried to constrain the presence of an absorption component by neutral hydrogen at the stellar radial velocity, which is the expected radial velocity of the stable gaseous disk. 

Introducing a second absorbing CS gas component at the stellar radial velocity (in addition to the ISM  and the redshifted component) did not improve the fit. As a consequence, we can only derive an upper limit on the CS hydrogen column density. We found that the column density must be $\log(N_{\mathrm{H}}/1\,\mathrm{cm}^2)_{\mathrm{RV}=0{\mathrm{km/s}}} \le 18.5$ at a 3\,$\sigma$ confidence level.

\section{Discussion}
\label{sec:discussion}

Our observations using the AVM technique in addition to a map of the airglow aimed at mitigating and correcting for the airglow contamination has allowed us to conduct the first measurements of the Lyman-$\alpha$ line toward $\beta$\,Pic. By fitting the line profile, we calculate the column density of neutral hydrogen in the line of sight toward the star to be $\log(N_{\mathrm{H}}/1\,\mathrm{cm}^2) = 18.6\pm0.1$ with a bulk radial velocity of about 40\,km/s relative to $\beta$\,Pic.

\begin{figure}
   \centering
   \includegraphics[width=\hsize]{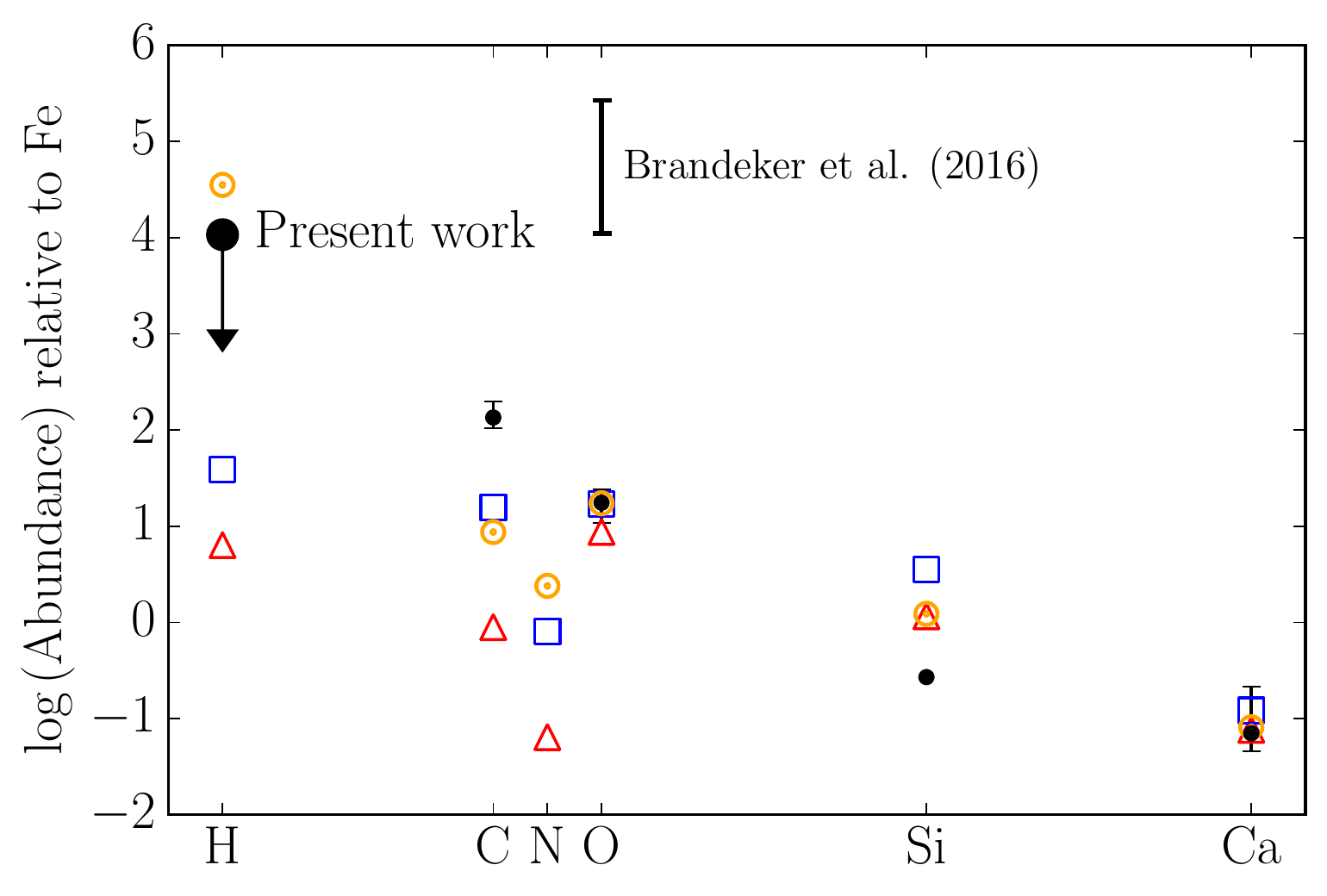}
      \caption{The abundances of the $\beta$\,Pic gas disk (black circles) compared to solar abundances (orange Sun symbols), CI Chondrites (red triangles) and Halley dust (blue squares). The abundances are given relative to iron (Fe) and the figure is adapted from Fig. 2 from \citep{roberge06}.
              }
         \label{fig:abundances}
   \end{figure}

The column density of all detected species in the circumstellar disk relative to iron are shown in Fig.\,\ref{fig:abundances}, together with the elemental abundances of CI Chondrites, Halley dust and the Sun. The figure shows that the derived abundance of hydrogen is higher than that of Halley dust and CI Chondrites, but lower than the solar abundance. The sub-solar abundance, being one order of magnitude lower than solar, indicates that the hydrogen cannot originate from the star which lets us exclude the possibility that the gas is the remnant of a protoplanetary disk. It is also highly unlikley that the hydrogen gas has been expelled by the star by radiation pressure and subsequently slowed down by another process as was suggested in \cite{lagrange98}.

The high column density of hydrogen compared to CI chondrites indicates that the gas was not predominantly created from dust grains in the disk, either by photon-stimulated desorption or by the vaporisation of colliding dust. The abundance of meteorites in the $\beta$\,Pic system may be different from that of our own Solar System. Despite this, if the gas were to originate from dust, we would expect to have detected an abundance of hydrogen orders of magnitude less than what we did.

The vaporisation of large bodies, such as exocomets, falling in towards $\beta$\,Pic may be supplying the detected hydrogen. Water is the main constituent in cometary atmospheres amongst the comets observed in the Solar System. They are known to have large envelopes of atomic hydrogen surrounding the nucleus which are created when the UV flux from the Sun photodissociates the cometary water into H and O via the sublimation at the nucleus or the surrounding grains \cite{mancuso15}. The photodissociation of H$_2$O by solar UV radiation produces H and O through the following reactions: 
\begin{align*}
\cee{H2O + $hv$ &-> OH + H\\
OH + $hv$ &-> O + H}
\end{align*}
If the photodissociation of cometary water was the primary process by which the observed hydrogen formed, one would expect the hydrogen and oxygen column densities to be similar, with the hydrogen column density being about twice that of oxygen.

Oxygen column density measurements by \citep{roberge06}, which are orders of magnitude smaller than that of hydrogen, would indicate that photodissociation of cometary water is unlikely to be the dominant hydrogen forming process, unless hydrogen and oxygen are subject to different removal mechanisms. For instance, we note that radiation pressure has a weak influence on hydrogen compared to the other elements including oxygen, and may cause a build up of hydrogen relative to oxygen with time. With similar production rates, an enhancement of hydrogen relative to oxygen could also be due to differences in the duration for which each element remains neutral. Despite the similar ionisation potentials of hydrogen and oxygen, if hydrogen remained neutral for longer, the amount of hydrogen could build up over time surpassing oxygen and carbon in abundance. Recent Herschel observations by \cite{brandeker16}, however, suggest the oxygen column density presented in \cite{roberge06} has been underestimated and that the actual value is $18 \lesssim \log(N_{\mathrm{OI}}/1\,\mathrm{cm}^2) \lesssim 20$. This supports the idea that the hydrogen gas can originate from exocometary water.

%For a definitive conclusion on the origin of the detected hydrogen in the $\beta$\,Pic environment, we need to better understand the gas dynamics and removal processes.

%{\bf{A large difference in the column density of hydrogen compared to oxygen \citep{roberge06} would suggests that the photodissociation of exocometary water is unlikely to be dominant hydrogen forming process. If it were the case, we would expect to measure a similar column density for hydrogen and oxygen, unless the lifetimes of these species differ significantly. Recent Herschel observations by \cite{brandeker16} suggest the oxygen column density presented in \cite{roberge06} has been underestimated and that the actual value is $18 \lesssim \log(N_{\mathrm{OI}}/1\,\mathrm{cm}^2) \lesssim 20$. This supports the idea that the hydrogen gas originates from exocomets.}}

Finally, the hydrogen detected here is seen moving toward the star, it is thus not the hydrogen corresponding to 
the stable circumstellar disk, as observed in other species or modelled by \cite{kral16}. 
As a consequence, and with only an upper limit for the hydrogen content of the stable gas disk, 
this new detection of hydrogen in $\beta$\,Pic raises new and interesting questions.

\section{Conclusions}
\label{sec:conclusions}

We calculate the column density of hydrogen in $\beta$\,Pic gas disk at zero radial velocity to be $\log(N_{\mathrm{H}}/1\,\mathrm{cm}^2) \ll 18.5$ and the column density of gas with a bulk velocity $\sim40$\,km/s to be $\log(N_{\mathrm{H}}/1\,\mathrm{cm}^2) = 18.6\pm0.1$. These two values are within the upper limits provided by radio observations of the 21\,cm hydrogen line. 

The high abundance of hydrogen relative to meteorites indicates that the gas is likely not produced from dust. On the other hand, the low abundance relative to the solar value excludes the possibility that the detected hydrogen could be a remnant of the protoplanetary disk or gas expelled by the star and driven out by radiation pressure. The observed hydrogen could be produced through dissociation of H$_2$O from evaporating exocomets transiting in front of the star on orbits with a narrow range of longitudes of periastron.

%The absorption at $\sim40$\,km/s, which is also seen in heavier species,

%In this scenario, the $\sim40$\,km/s radial velocity of the redshifted peak of the absorption would be, as for the redshifted absorption features seen in heavier species, explained by evaporating exocomets transiting in front of the star on orbits with a narrow range of longitudes of periastron.

%Nonetheless, with a hydrogen abundance almost three orders of magnitude higher than oxygen this scenario is challenged, unless the oxygen is more efficiently removed compared to hydrogen. 

\begin{acknowledgements}
P.A.W, A.L.E and A.V-M all acknowledge the support of the French Agence Nationale de la Recherche (ANR), under program ANR-12-BS05-0012 "Exo-Atmos". Fig.\,\ref{fig:mcmc} was created using \texttt{corner.py} written by \cite{dan_foreman_mackey_2016_53155}. The authors would like to acknowledge the anonymous referee for their comments.
\end{acknowledgements}

\bibliography{cos}

\clearpage
\appendix

\section{Flux loss with the AVM technique}
\label{sec:appendix}
In this appendix we develop a mathematical model for the flux loss induced by moving a target along the dispersion axis, executed on purpose to increase the separation between the target spectrum and the airglow spectrum. Separating the spectra in wavelength is an ideal way of recovering more of the target spectrum and limiting the contamination by airglow.

\subsection{A mathematical model for flux loss.}
An increase in the off-axis target distance along the dispersion axis effectively shifts the Ly-$\alpha$ emission feature in wavelength relative to the Ly-$\alpha$ airglow. This is because the light from the shifted target lands on a different part of the detector whilst the airglow emission continues to fill the entire aperture and does not change its position on the detector when the aperture is moved. Shifting the aperture in the positive dispersion direction shifts the target spectrum redward, revealing the red side of the Ly-$\alpha$ line and vice versa. An increase in the separation between the target spectrum and airglow spectrum comes at a loss in flux as the aperture increasingly blocks more flux. We calculate the loss in flux, $F(\rho)$, as a function of distance from the centre of the aperture, $\rho$, along the dispersion axis (along $x$) assuming a Gaussian PSF,

   \begin{figure}
   \centering
   \includegraphics[width=4cm]{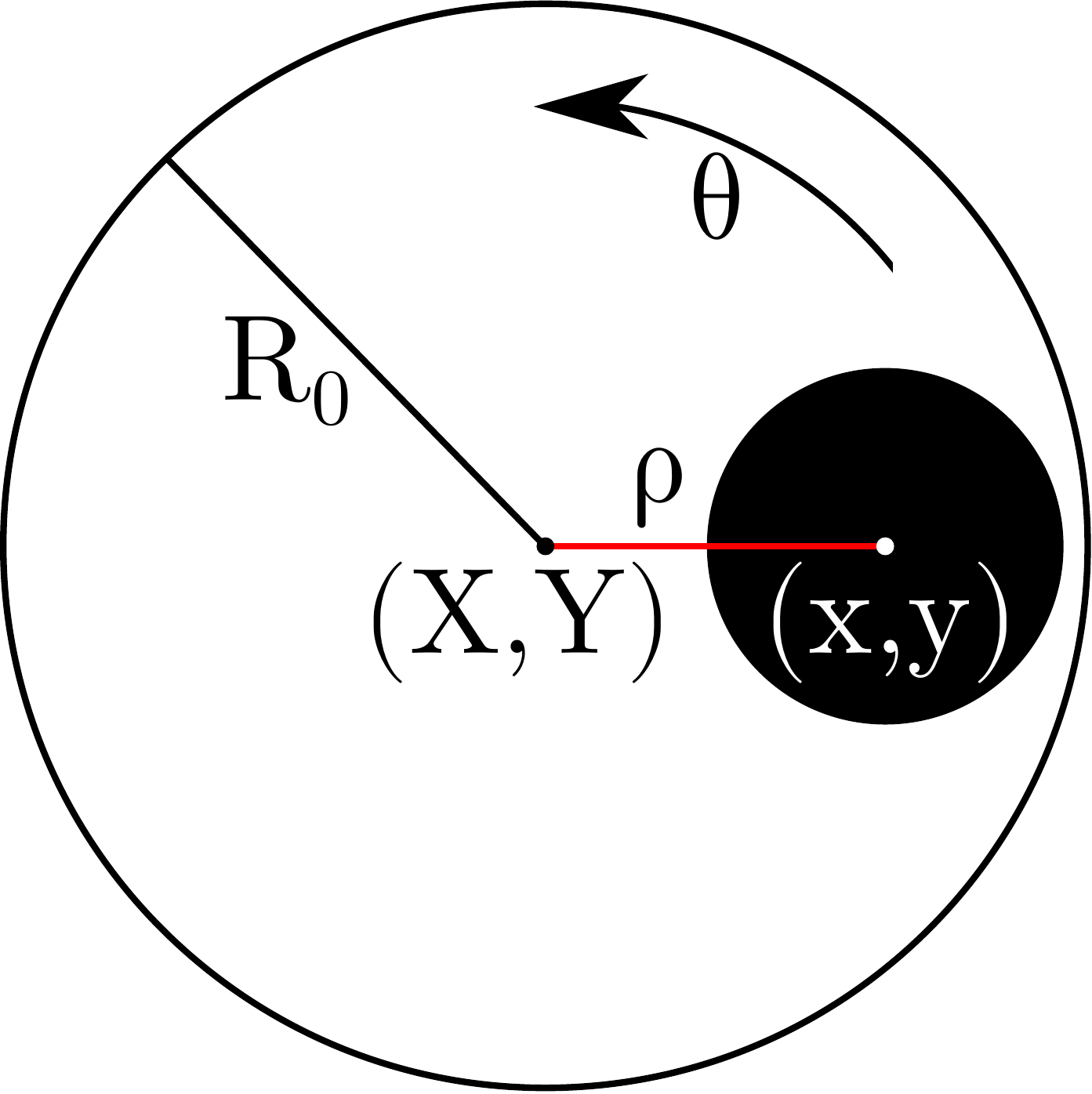}
      \caption{Graphical representation of the PSF (black) shifted along the dispersion axis by an amount $\rho$ in an aperture of radius $R_0$.
              }
         \label{fig:graphic}
   \end{figure}

   \begin{figure}
   \centering
   \includegraphics[width=\hsize]{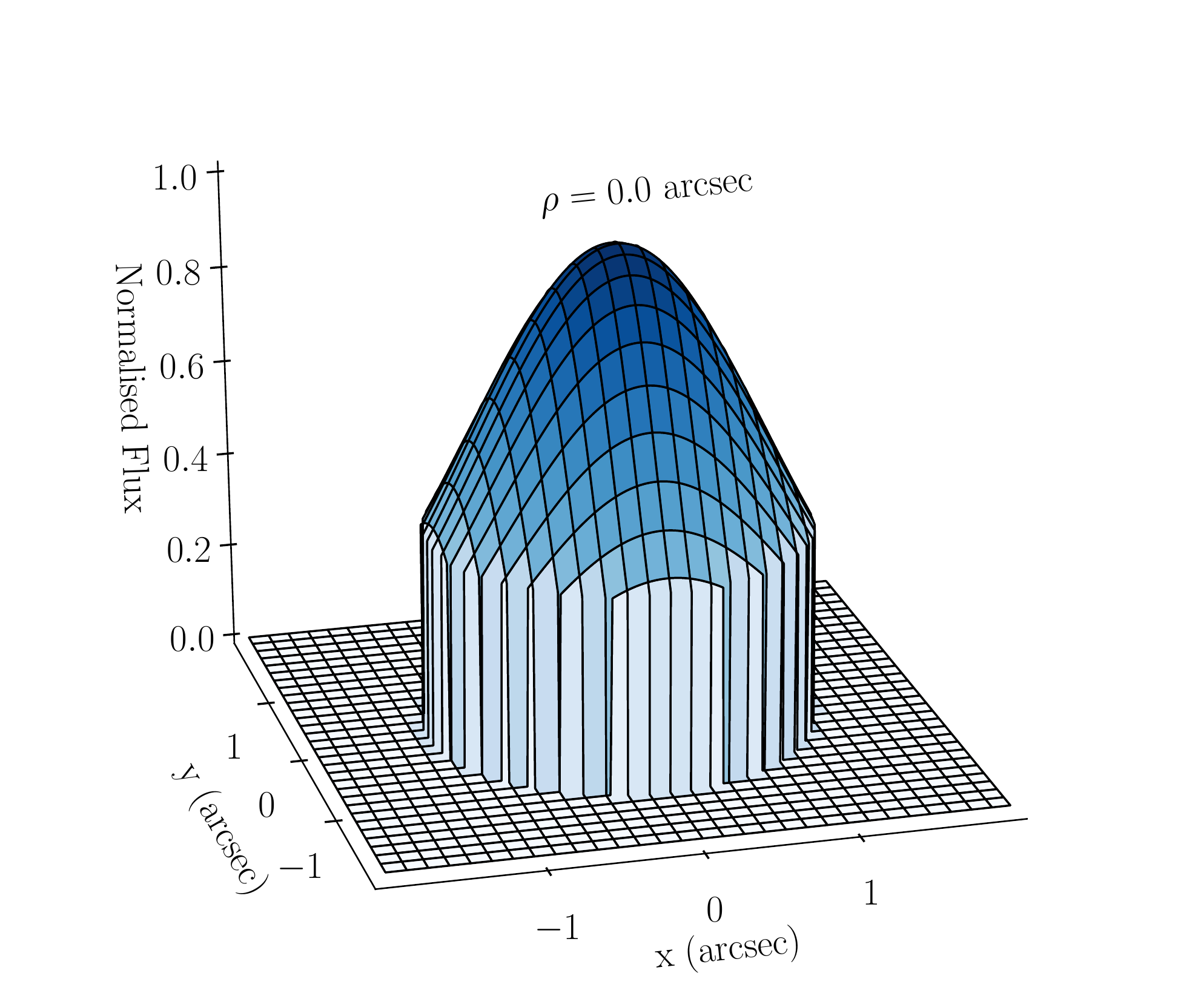}
   \includegraphics[width=\hsize]{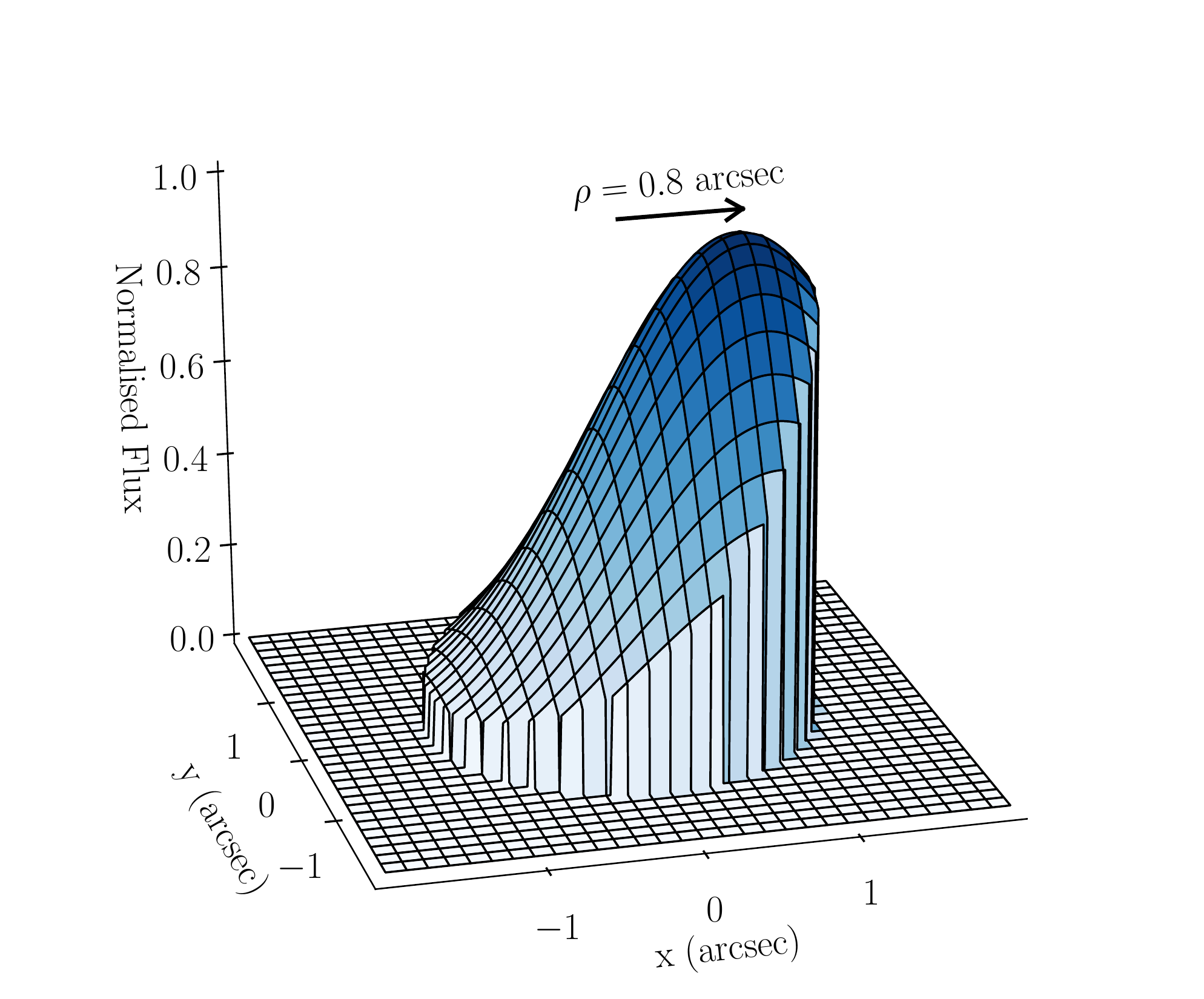}
   \includegraphics[width=\hsize]{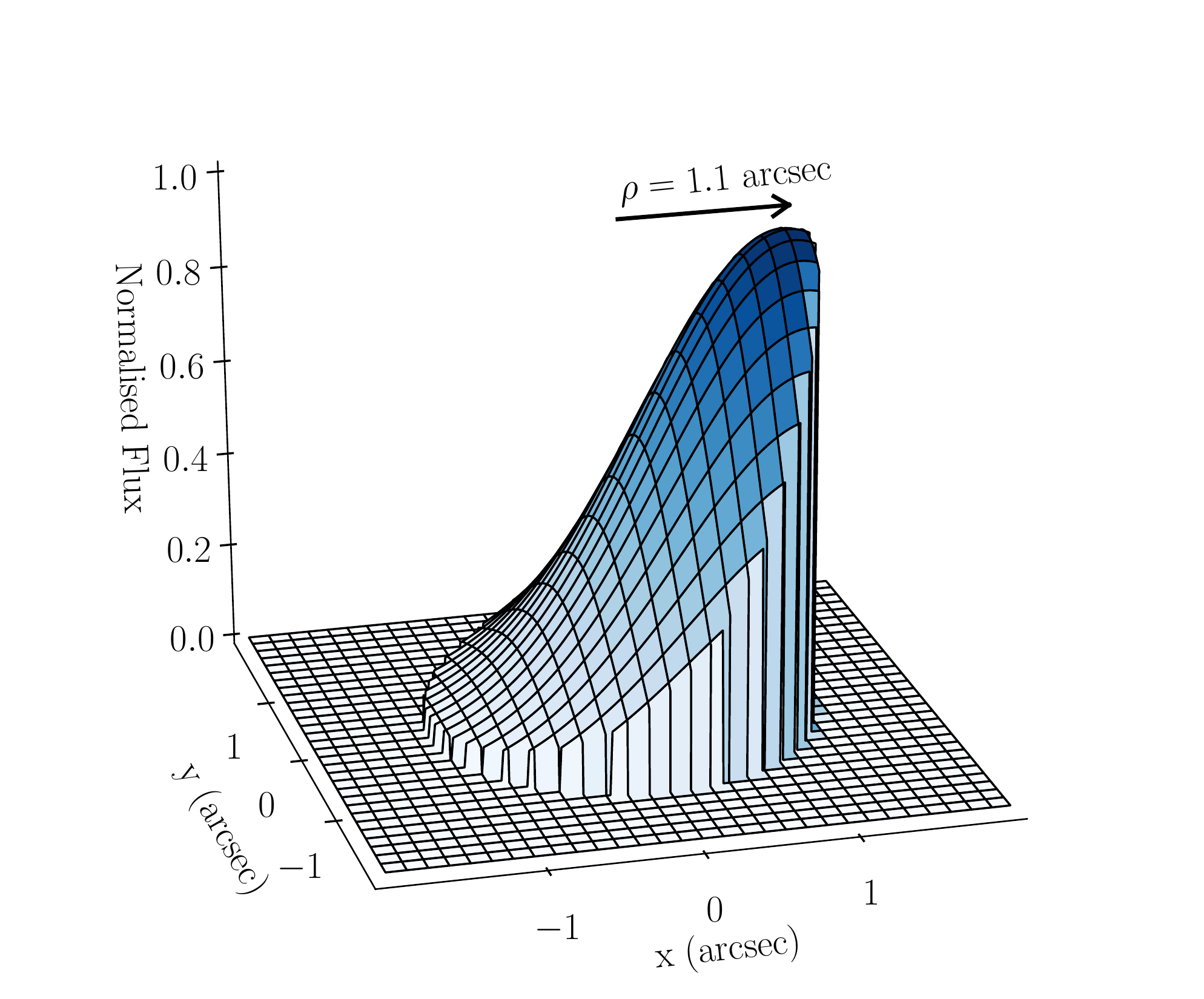}
      \caption{The normalised flux coming through a circular aperture. {\it{Top:}} The aperture is centred on the target, $F(\rho=0\arcsec)\equiv1$. {\it{Middle:}} The target is 0.8\arcsec offset along the dispersion axis, $F(0.8\arcsec) =0.72$. {\it{Bottom:}} The target is 1.1\arcsec offset along the dispersion axis, $F(1.1\arcsec) =0.58$.
              }
         \label{fig:shift}
   \end{figure}

   \begin{figure}
   \centering
   \includegraphics[width=\hsize]{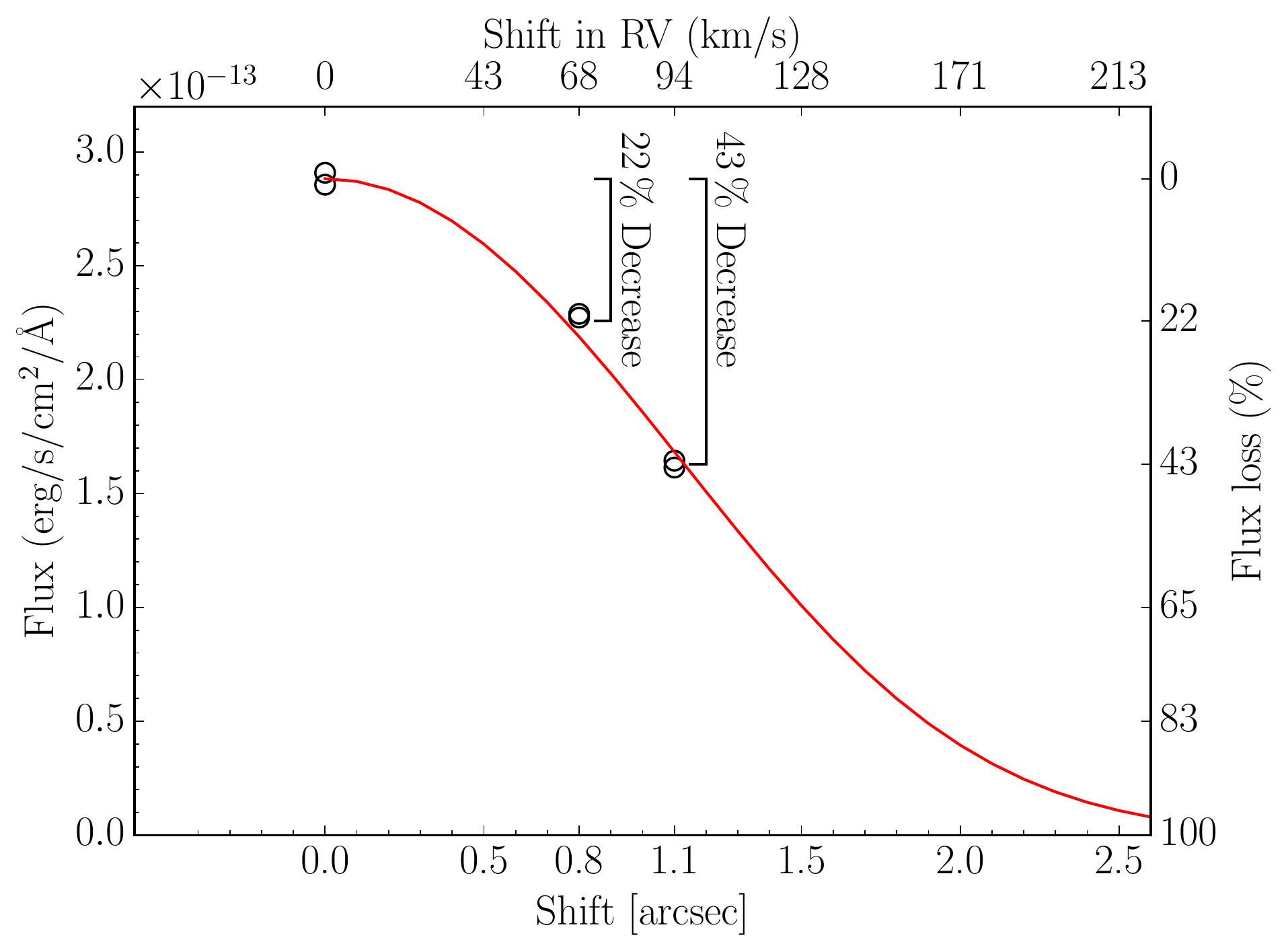}
      \caption{A positional shift of the aperture away from the central position results in a flux loss as indicated with the red line. The black circles represent the measured flux level for the 26 December 2015 and 30 January 2016 measurements when shifted by 0.0, +0.8 and +1.1\,$\arcsec$. A shift $\pm0.8$\,arcsecond away from the central position results in a 68\,km/s RV shift of the Ly-$\alpha$ line and a 22\% decrease in flux. The 1.1\,arcsecond shift resulted in a 94\,km/s shift in the Ly-$\alpha$ line and a 43\% decrease in flux.
              }
         \label{fig:flux_loss}
   \end{figure}

\begin{equation}
\label{eq:1}
F(\rho) = F_0 \int_{-\infty}^{\infty} \int_{-\infty}^{\infty} e^{\frac{-(x^2+y^2)}{2 \sigma^2}}\mathrm{d}x \mathrm{d}y \Theta (R_0-R)
\end{equation}

\noindent where $x$ and $y$ are the Cartesian coordinates relative to the central part of the PSF and $\Theta$ the aperture. The aperture, $\Theta$, is modelled as a Heaviside step function where,

 \begin{equation}
   \Theta(R_0-R) = \left\{
     \begin{array}{lr}
       0 & : R > R_0 \\
       1 & : R < R_0.
     \end{array}
   \right.
\end{equation} 

\noindent The dispersion $\sigma$ is calculated from the FWHM by the relation

\begin{equation}
\mathrm{FWHM} = 2\sqrt{2\ln2}\sigma.
\end{equation}

\noindent By making a change from the Cartesian coordinate system to one using polar coordinates the integral limits are no longer unbounded and we are able to derive an analytical equation with finite integral limits which can more easily be graphically represented. By choosing the aperture as the reference frame instead of the PSF allows us to do the substitutions $r = X - \rho$ and $\mathrm{d}x \mathrm{d}y$ = $\mathrm{d}X \mathrm{d}Y$ with Eq.\,\ref{eq:1} being expressed as,

\begin{equation}
F(\rho) = F_0 \int_{-\infty}^{\infty} \int_{-\infty}^{\infty} e^{\frac{-(X^2+Y^2)}{2 \sigma^2}} e^{\frac{(2X\rho - \rho^2)}{2\sigma^2}} \mathrm{d}X \mathrm{d}Y \Theta (R-R_0).
\end{equation}

\noindent The change to polar coordinates allows us to express $\mathrm{d}X \mathrm{d}Y$ as $2\pi R \mathrm{d}\theta \mathrm{d}R$ and $X = R\cos \theta$, which finally gives us

\begin{equation}
F(\rho) = F_0e^{\frac{-\rho^2}{2\sigma^2}}\int_0^{2\pi}\int_0^{R_0}e^{\frac{2\rho R \cos \theta - R^2}{2\sigma^2}} R \mathrm{d}R \mathrm{d}\theta
\end{equation}

\noindent which we represent graphically in Fig.\,\ref{fig:flux_loss}.

\subsection{Comparison with observations}
We find that the measured loss in flux is well modelled with a FWHM of $\sim2.3$\arcsec. Previous measurements of cross-dispersion profile widths with the G130M grating combined with the 1291\,\AA\ central wavelength setting have shown FWHM widths of $\sim1.4-1.6$\arcsec\ for the two first lifetime positions of the instrument and assuming a Gaussian fit \citep{debes16}. We attribute this increased FWHM to the mid-frequency wavefront errors caused by the non-Gaussian wings they introduce and the uncorrected astigmatism in the cross-dispersion direction. The change in lifetime position for the 2015 and 2016 observations presented in this paper may also differ from previously measured lifetime positions. The loss in flux as a function of $\rho$ is shown in Fig.\,\ref{fig:flux_loss} together with the calculated shift (top axis). We find that a 50\% loss in flux relative to $\rho=0$\,\arcsec occurs when $\rho\sim1.2$\,\arcsec.

\end{document}